\documentclass[
  prb,
  twocolumn,
  final,
  floats,
  floatfix,
  amsmath,
  amssymb,
  nofootinbib,
  eqsecnum,
  byrevtex,
]{revtex4}

\usepackage{graphicx}

\def\rr{\mathbf{r}}
\def\ud{\mathrm{d}}
\def\peptide{\texttt{Ace}\texttt{-}\texttt{GGPGGG}\texttt{-}\texttt{Nme }}

\begin{document}
\title{
Adaptively Biased Molecular Dynamics for Free Energy Calculations
}
\author{Volodymyr Babin}
\author{Christopher Roland}
\author{Celeste Sagui}\email{sagui@ncsu.edu}
\affiliation{
Center for High Performance Simulations (CHiPS) and
Department of Physics, North Carolina State University, Raleigh,
NC 27695
}
\date{January 16, 2008}
\begin{abstract}
We present an \texttt{A}daptively \texttt{B}iased \texttt{M}olecular
\texttt{D}ynamics (\texttt{ABMD}) method for the computation of the free energy
surface of a reaction coordinate using non-equilibrium dynamics.
The \texttt{ABMD} method belongs to the general category of umbrella sampling
methods with an evolving biasing potential, and is inspired 
by the metadynamics method. The \texttt{ABMD} method has several useful
features, including a small number of control parameters, and an $O(t)$
numerical cost with molecular dynamics
time $t$. The \texttt{ABMD} method naturally allows for extensions based
on \textit{multiple walkers} and \textit{replica exchange}, where different
replicas can have different temperatures and/or collective variables.
This is beneficial not only in terms of the speed and accuracy of
a calculation, but also in terms of the amount
of useful information that may be obtained from a given simulation.
The workings of the \texttt{ABMD} method are illustrated via a study of the
folding of the \peptide peptide in a gaseous and solvated environment.
\end{abstract}
\maketitle
%
%
\section{\label{sec:introduction}
Introduction
}
When investigating the equilibrium properties of a complex polyatomic
system, it is customary to identify a suitable \textit{reaction coordinate}
$\sigma(\rr_1,\dots,\rr_N):\mathbb{R}^{3N}\mapsto\mathbb{Q}$ that
maps atomic positions $\rr_1,\dots,\rr_N$ onto the points of some manifold
$\mathbb{Q}$, and then to study its equilibrium probability density:
\begin{equation}
  \nonumber
  p(\xi) = \big<\delta\left[\xi - \sigma(\rr_1,\dots,\rr_N)\right]\big>,\;
  \xi\in\mathbb{Q}
\end{equation}
(angular brackets denote an ensemble average). The density $p(\xi)$
provides information about the relative stability of states corresponding to
different values of $\xi$ along with useful insights into the transitional
kinetics between various stable states. In practice, the  Landau free
energy\cite{Frenkel} 
\begin{equation}
  \nonumber
  f(\xi) = -k_BT\ln p(\xi),
\end{equation}
is typically preferred over $p(\xi)$, 
because it tends to be more intuitive. Either $p(\xi)$ or $f(\xi)$ is
said to provide a coarse-grained description of the system --- in
terms of $\xi$ alone --- with the rest of the degrees of freedom
of the original system integrated out. Quite naturally, the reaction
coordinate (often also referred to as \textit{collective variable} or
\textit{order parameter}) is typically chosen to represent the
slowest degrees of freedom of the original system, although this is
not formally required.

In the past few years, several methods targeting the computation
of $f(\xi)$ using non-equilibrium dynamics have become popular. 
First methods that introduced a time evolving potential to bias
the original potential energy were
the the \texttt{L}ocal \texttt{E}levation
\texttt{M}ethod (\texttt{LEM})\cite{Huber_T_94},
by Huber, Torda
and van Gunsteren in the \texttt{MD} context and the Wang-Landau approach
in \texttt{MC} one\cite{Wang_F_01}. More recent approaches include the
adaptive-force bias method\cite{Darve_E_01}, and the
non-equilibrium metadynamics\cite{Laio_A_02,Iannuzzi_M_03} method.
These methods all estimate the free energy of the reaction
coordinate from an ``evolving'' ensemble of realizations%
\cite{Lelievre_T_2007,Bussi_G_06}, and use
that estimate to bias the system dynamics, so as to flatten the effective free
energy surface. Collectively, they can all be considered as umbrella sampling
methods, with an evolving potential. In the long time limit, the biasing force
is expected to compensate for the free energy gradient, so that the biasing
potential eventually reproduces the free energy surface.

In this work, we present an \texttt{A}daptively \texttt{B}iased
\texttt{M}olecular \texttt{D}ynamics (\texttt{ABMD}) method whose
implementation is particularly efficient and suited for free energy calculations.
The method has an $O(t)$ scaling with molecular dynamics time t and is
characterized by only two control parameters. In addition,
the method allows for extensions based on {\it multiple walkers} and
{\it replica exchange} for both temperature and/or the collective variables.
The \texttt{ABMD} method has been implemented
in the \texttt{AMBER} software package\cite{Amber9}, and is to be distributed
freely.

Before discussing \texttt{ABMD}, it is helpful to review
the salient features of the metadynamics (\texttt{MTD}) method.
Essentially, the \texttt{MTD} method is built upon the \texttt{LEM} method
by exploiting Car-Parrinello dynamics: the phase space of the
system is extended to include additional dynamical degrees of freedom
harmonically coupled to the collective variable. These additional degrees
of freedom are assumed to have masses associated with them, and evolve
in time according to Newton's laws. The masses are supposed to be large
enough, so that the dynamics of these extra-variables is driven by the
free energy gradient. Their trajectory is then used to construct
a history-dependent biasing potential by means of placing many small
Gaussians along the trajectory. When combined with Car-Parrinello
\textit{ab-initio} dynamics, \texttt{MTD} has been successfully used to
explore complex reaction pathways involving several energy
barriers\cite{Ensing_B_04,%
Churakov_S_04,Gervasio_F_05,Ceccarelli_M_04,Iannuzzi_M_04,Stirling_A_04%
,Asciutto_E_05,Lee_J_06,Ikeda_T_05}.

While \texttt{MTD} continues to be used successfully, there
are several known limitations associated with the initial implementation of
the method, which provided the motivation for the development of the
\texttt{ABMD} method. First, in order
to calculate reliable free energies with a \textit{controllable accuracy},
long runs are needed, especially for the ``corrective'' follow-up
at equilibrium\cite{Babin_V_2006}.
This is especially true for biomolecular systems, which typically are
characterized by many degrees of freedom and non-negligible entropy
contributions to the free energies. Long runs, however, may be
precluded by the \texttt{MTD} method, because of its unfavorable
scaling with molecular dynamics \texttt{MD} time $t$.
While one can readily speed up the original \texttt{MTD} 
method using such tricks as truncated Gaussians and kd-trees\cite{Babin_V_2006},
the bottleneck there is  the explicit calculation of the history-dependent
potential.
Since at every \texttt{MD} step, Gaussians from all previous time steps need to
be added, the number of Gaussians grows linearly with $t$. The numerical cost of
\texttt{MTD} therefore grows as $O(t^2)$ which, in some cases, may prove itself to
be prohibitively expensive, especially when long runs are needed.
Another undesirable feature is that the \texttt{MTD}
method 
(at least in its original implementation)
is characterized by a relatively large number of parameters
({\it e.g.}, the masses and spring constants associated with the collective
variable, the characteristics of the Gaussians to be added, multiple control
parameters, etc), all of which
influence the dynamics in an entangled and non-transparent way. A successful
\texttt{MTD} run often requires a careful balancing of these parameters, which
is especially nontrivial for multidimensional collective variables.
More recent implementations of \texttt{MTD}~\cite{Laio_A_05} have reduced
the number of parameters. 
As will be discussed, the \texttt{ABMD} method 
is characterized by only {\it two} control parameters
and scales as $O(t)$ with simulation time.
%
%
\section{\label{sec:method}
The Adaptively Biased Molecular Dynamics Method.
}
The \texttt{ABMD} method is formulated in terms of the following
set of equations:
\begin{equation}
  \nonumber
  m_a\frac{\mathrm{d}^2\rr_a}{\mathrm{d}t^2} = \mathbf{F}_a
  - \frac{\partial}{\partial\rr_a}
    U\big[t|\sigma\left(\rr_1,\dots,\rr_N\right)\big],
\end{equation}
\begin{equation}
  \nonumber
  \frac{\partial U(t|\xi)}{\partial t} = \frac{k_B T}{\tau_F}
    G\big[\xi - \sigma\left(\rr_1,\dots,\rr_N\right)\big],
\end{equation}
where the first set represents Newton's equations that govern ordinary
\texttt{MD} (temperature and pressure regulation terms are
not shown) augmented with the additional force coming from the time-dependent
biasing potential $U(t|\xi)$ (with $U(t=0|\xi) = 0$), whose time evolution
is given by the second equation. In the following, we refer to $\tau_F$ as
\textit{flooding timescale} and to $G(\xi)$ as \textit{kernel}
(in analogy to the \textit{kernel density estimator} widely used in
statistics\cite{Silverman_Density_Estimation}).
The kernel is supposed to be positive definite ($G(\xi)>0$) and symmetric
($G(-\xi)=G(\xi)$). It can be perceived as a smoothed Dirac delta function.
For large enough $\tau_F$ and small enough width of the kernel,
the biasing potential $U(t|\xi)$ converges towards $-f(\xi)$ as
$t\to\infty$\cite{Lelievre_T_2007,Bussi_G_06}.

Our numerical implementation of the \texttt{ABMD} method involves
the following. We 
stick with $\mathbb{Q} = \mathbb{Q}_1\times\dots\times\mathbb{Q}_D$
where $\mathbb{Q}_k$ is either $[a,b]\in\mathbb{R}^1$ or a one-dimensional
torus, and use cubic B-splines (or products of thereof for $D>1$) to
discretize $U(t|\xi)$ in $\mathbb{Q}$:
\begin{equation}
  \nonumber
  U(t|\xi) = \sum\limits_{m\in\mathbb{Z}^D} U_m(t)B(\xi/\Delta\xi - m),
\end{equation}
\begin{equation}
  \nonumber
  B(\xi) = \left\{
    \begin{array}{ll}
      (2 - |\xi|)^3/6, & 1\leq |\xi| < 2, \\
      \xi^2(|\xi| - 2)/2 + 2/3, & 0\leq |\xi| < 1, \\
      0, & \mathrm{otherwise}.
    \end{array}
  \right.
\end{equation}
We use the biweight kernel\cite{Silverman_Density_Estimation} for $G(\xi)$:
\begin{equation}
  \nonumber
  G(\xi) = \frac{48}{41}\left\{
    \begin{array}{ll}
      \left(1 - \left.\xi^2\right/4\right)^2, & -2 \leq \xi \leq 2 \\
      0, & \mathrm{otherwise}
    \end{array}
  \right.,
\end{equation}
and an Euler-like discretization scheme for the time evolution
of the biasing potential:
\begin{equation}
  \nonumber
  U_m(t + \Delta t) = U_m(t)
  + \Delta t\frac{k_BT}{\tau_F}G\big[\sigma/\Delta\xi - m\big],
\end{equation}
where $\sigma = \sigma(\rr_1,\dots,\rr_N)$ is at time $t$. Note that
this time discretization may be readily improved.
This, however, is not really important here, since the goal is not
to recover the solution of the \texttt{ABMD} equations \textit{per se}, but
rather to flatten $U(t|\xi) + f(\xi)$ in the $t\to\infty$ limit. Note also,
that the numerical cost of evaluation of the time-dependent
potential is constant over time, and so \texttt{ABMD} scales
trivially as $O(t)$, which is computationally quite favorable.
The storage requirements of the \texttt{ABMD} are also quite reasonable,
especially if sparse arrays are used for $U_m$.
In addition, it is characterized by only {\it two}
control parameters: the flooding timescale $\tau_F$ and the kernel width
$4\Delta\xi$.

\texttt{ABMD} admits two important extensions. 
The first is identical in spirit to the \textit{multiple walkers
metadynamics}\cite{Lelievre_T_2007,Raiteri_P_2006}. It amounts to carrying
out several different \texttt{MD} simulations biased by the same
$U(t|\xi)$, which evolves via: 
\begin{equation}
  \nonumber
  \frac{\partial U(t|\xi)}{\partial t} = \frac{k_B T}{\tau_F}
    \sum\limits_{\alpha}%
      G\big[\xi - \sigma\left(\rr_1^{\alpha},\dots,\rr_N^\alpha\right)\big]\;,
\end{equation}
where $\alpha$ labels different \texttt{MD} trajectories.
A second extension is to gather several different \texttt{MD}
trajectories, each bearing its own biasing potential and, if desired,
its own  distinct collective variable, into a
generalized ensemble for ``replica exchange'' with modified ``exchange'' 
rules\cite{Sugita_Y_2000,PTMetaD,BiasExchange}.
Both extensions are advantageous and lead to a more uniform
flattening of $U(t|\xi) + f(\xi)$ in $\mathbb{Q}$. This enhanced
convergence to $f(\xi)$ is due to the improved
sampling of the ``evolving'' canonical distribution.
\begin{figure}[t!]
\includegraphics[width=\linewidth,clip=true]{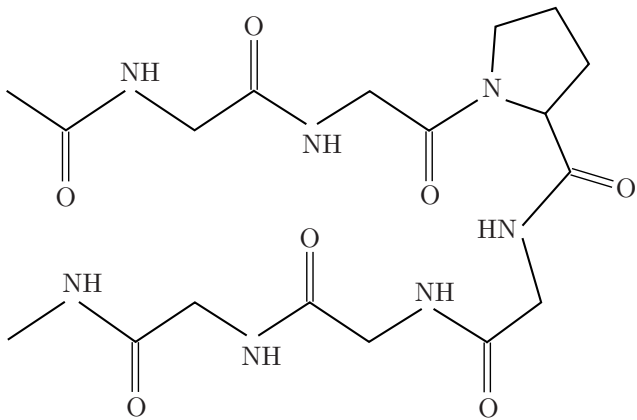}
\caption{\label{fig:1}%
The {\peptide} peptide in a $\beta$-hairpin conformation (sketch).
}
\end{figure}

We have implemented the \texttt{ABMD} method in the \texttt{AMBER}
package\cite{Amber9}, with
support for both replica exchange and multiple-walkers. In pure
``parallel tempering'' replica exchange (same collective variable
in all replicas), $N_r$ replicas
are simulated at different temperatures
$T_n$, $n=1,\dots,N_r$. Each replica has its own biasing potential
$U^n(t|\xi)$, $n=1,\dots,N_r$, that evolves according to its dynamical equation. 
Exchanges between neighboring replicas are attempted at a prescribed
rate, with an exchange probability given by\cite{Sugita_Y_2000}:
\begin{equation}
  \label{eq:exchange1}
  w(m|n) = \left\{%
    \begin{array}{ll}
    1, & \Delta\leq 0,\\
    \exp (-\Delta), & \Delta > 0,
    \end{array}
  \right.
\end{equation}
\begin{eqnarray}
  \label{eq:exchange2}
  \Delta &=& \left(\frac{1}{k_BT_n} - \frac{1}{k_BT_m}\right)
           \Big(E_p^m - E_p^n\Big)\\
    \nonumber
    &+& \frac{1}{k_BT_m}\Big[U^m(\xi^n) - U^m(\xi^m)\Big]\\
    \nonumber
    &-& \frac{1}{k_BT_n}\Big[U^n(\xi^n) - U^n(\xi^m)\Big],
\end{eqnarray}
where $E_p$ denotes the atomic potential energy.
The biasing potentials are temperature-bound and converge in the $t\to\infty$
limit to the free energies at their respective temperatures.

We have also implemented a more \textit{general} replica exchange scheme,
where different replicas can have different collective variables and/or 
temperatures, and can experience  either an evolving or a static biasing potential
(the latter obviously includes the case of $U^n(t|\xi)=0$).
Exchanges between random pairs of replicas are then tried at a prescribed rate.
This method is simply a generalization~\cite{BiasExchange} of the ``Hamiltonian
replica exchange'' method described in Ref.\onlinecite{Sugita_Y_2000}, and
reduces to it when all biasing potentials are static.
The big advantage here is that, by using replicas with different collective
variables, it is possible to obtain several \textit{one-} or
\textit{two-}dimensional
projections of the free energy surface for the corresponding variables.
This is very useful because it not only increases the amount of information
that can be gathered from a given simulation,  but it also allows for
previously obtained information for a collective variable to be used
to compute the free energy associated with different variables.
For instance, suppose that in the course of a simulation it becomes apparent
that one wishes to address additional questions involving different collective
variables. Instead of starting from ``scratch'', one
can re-use the already obtained biasing potentials
and thereby greatly accelerate the free energy calculation for the new variables.
It is also worth noting that for replicas running at the same temperature, the
exchange probability does not depend on the atomic potential energies
(Eqs.~(\ref{eq:exchange1})-(\ref{eq:exchange2}) above). This implies that
the number of replicas needed to maintain acceptable exchange rates can be
made independent of the solvent degrees of freedom, provided that one is interested in
the properties of the solute only  (so that the collective variables
do not depend explicitly on the atomic coordinates of the solvent)
and that the structure is adequately solvated. This can be exploited to
sample a solute with a minimum amount of solvent, and to accelerate
the averaging over the solvent degrees of freedom. 
These last two applications of the \textit{general} replica exchange method are illustrated
in the next section.
\begin{figure}[t!]
\includegraphics[width=\linewidth,clip=true]{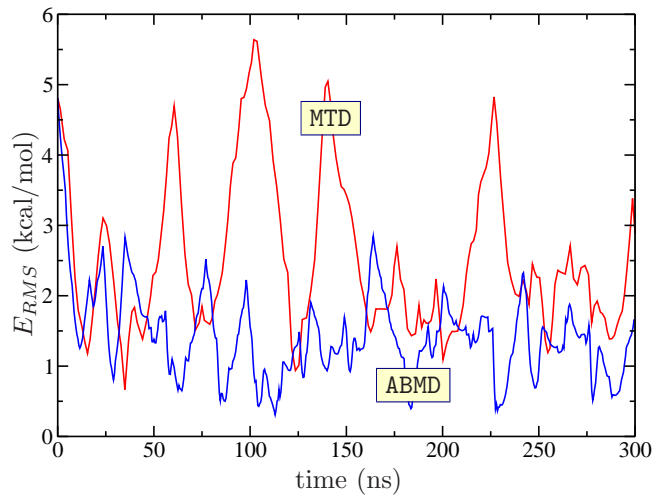}
\caption{\label{fig:2}%
RMS error of the free energy over $3.3\text{\AA}<R_g<6.3\text{\AA}$ at
$T=300K$ for the \texttt{ABMD} (blue) and reference \texttt{MTD}
(red) simulations.
}
\end{figure}
%
%
\section{\label{sec:output}
Case study: a short peptide.
}
To illustrate the method, we have simulated the hydrophobic \peptide peptide
(sketched in Fig.\ref{fig:1}) in the gas phase  and in a solvated environment, 
using cyclohexane as the explicit solvent.
The free energy of this peptide at $T=300\,\mathrm{K}$ in gas phase has previously been
investigated with the \texttt{MTD} method\cite{Babin_V_2006}, and is
characterized by a ``double-well'' structure (see 
Fig.\ref{fig:8}), with the wells corresponding to the peptide in
a ``globular'' (left minimum in the Fig.\ref{fig:8}) and
a $\beta$-hairpin (right minimum in the Fig.\ref{fig:8})
folded conformation, respectively. While simple
enough, the molecule possesses all the typical features of larger peptide
systems usually studied with biomolecular simulations. Simulation
parameters are as in a previous study\cite{Babin_V_2006}: the atoms are
described by the 1999 version of the Cornell \textit{et al.} force 
field\cite{Cornell_W_95}, with no cutoff for the non-bonded interactions.
The Berendsen thermostat is chosen with $\tau_{tp}=1\mathrm{fs}$ for
temperature control. The \texttt{MD} time step ($\Delta t$)
is 1 fs for the parallel tempering simulations, and 2 fs otherwise.
\begin{figure}[t!]
\includegraphics[width=\linewidth,clip=true]{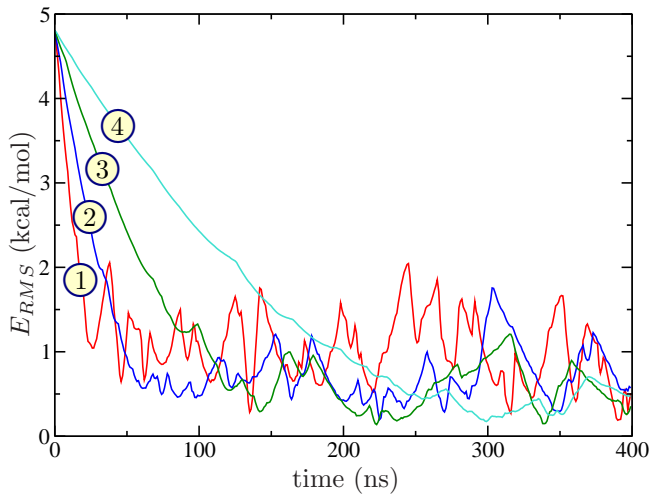}
\caption{\label{fig:3}%
RMS error of the free energy over $3.3\text{\AA}<R_g<6.3\text{\AA}$ for
\texttt{ABMD} simulations at $T=300K$ with $\tau_F = 180\;\mathrm{ps}$ (1),
$360\;\mathrm{ps}$ (2), $720\;\mathrm{ps}$ (3) and
$1440\;\mathrm{ps}$ (4).
}
\end{figure}

The radius of gyration of the heavy atoms was chosen to be
the collective variable:
\begin{equation}
  \label{eq:rg}
  R_g = \sum_{a}\frac{m_a}{m_{\Sigma}}\left(\rr_a
                         - \mathbf{R}_{\Sigma}\right)^2\,.
\end{equation}
Here, $\mathbf{R}_{\Sigma} = \sum_{a}\left(m_a/m_{\Sigma}\right)\rr_a$ is the
center of mass, with $m_{\Sigma} = \sum_{a}m_a$, and the sum runs over all
atoms except hydrogen. The initial configuration 
is the fully unfolded peptide. A reference free energy profile, whose
error\cite{Babin_V_2006} in the region of interest is
less than 0.15 kcal/mol, was computed for benchmarking purposes
(see Appendix~\ref{ap:reference} for details). As a measure of the RMS free
energy error, the following construction was used:
\begin{equation}
  \nonumber
  E_{RMS} = \sqrt{\frac{1}{b - a}\int\limits_{a}^{b}%
    \ud\xi\Big(f_1(\xi) - f_2(\xi) - \Delta\Big)^2},
\end{equation}
where
\begin{equation}
  \nonumber
  \Delta = \frac{1}{b - a}\int\limits_{a}^{b}%
  \ud\xi\Big(f_1(\xi) - f_2(\xi)\Big),
\end{equation}
accounts for the arbitrary additive constants in the free
energies $f_{1,2}(\xi)$. Here $a=3.3${\AA} and $b=6.3${\AA}, which correspond
to the physical region of interest (see Fig.\ref{fig:8}).
\begin{figure}
\includegraphics[width=\linewidth,clip=true]{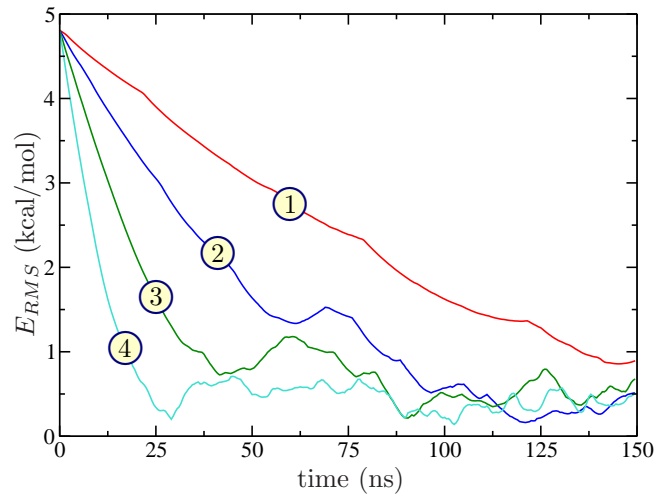}
\caption{\label{fig:4}%
RMS error of the free energy over 3.3{\AA}$<R_g<$6.3{\AA} for multiple
walkers \texttt{ABMD} simulations with
$\tau_F = 1\,\mathrm{ns}$ using one (1), two (2),
four (4) and eight (8) trajectories at $T = 300$K.
}
\end{figure}

Figure~\ref{fig:2} presents the time dependence of the
RMS free energy error for the \texttt{ABMD} and reference \texttt{MTD}
run\cite{Babin_V_2006}. Both simulations 
have exactly the same kernel width $4\Delta\xi = 0.25\;${\AA}, and
flooding timescale $\tau_F=90\;\mathrm{ps}$ that corresponds to the
\textit{a posteriori} hills acceptance rate reported in
Ref.\onlinecite{Babin_V_2006}. It is evident
that the \texttt{AMBD} run is more accurate than the corresponding \texttt{MTD} run.
The \texttt{ABMD} method owes its better convergence to the smoother time
evolution of the biasing potential and accurate discretization
in $\mathbb{Q}$. The amount of memory used by the \texttt{ABMD} simulation
to store $U_m$ values was only $\approx 200\times 8$ bytes
(considering double precision) for roughly $10^8$ tiny ``hills''
that were accumulated by the end of the run. The reference
\texttt{MTD} simulation with merely $5\times 10^3$ Gaussians
required roughly 25 times more memory for the biasing potential
(with only the  positions of the Gaussians stored explicitly). 
One can expect, that \texttt{ABMD} will be even more economical when it
comes to dealing with multidimensional collective variables, provided
that sparse arrays are used for $U_m$ with only non-zero elements
being stored explicitly.

Although an \textit{a priori} error estimate for this type of non-equilibrium
simulation is really not feasible, it is expected that the error should
decrease for increased $\tau_F$. This point is illustrated in
Fig.\ref{fig:3}, which shows the error for increasing values of $\tau_F$.

\begin{figure}
\includegraphics[width=\linewidth,clip=true]{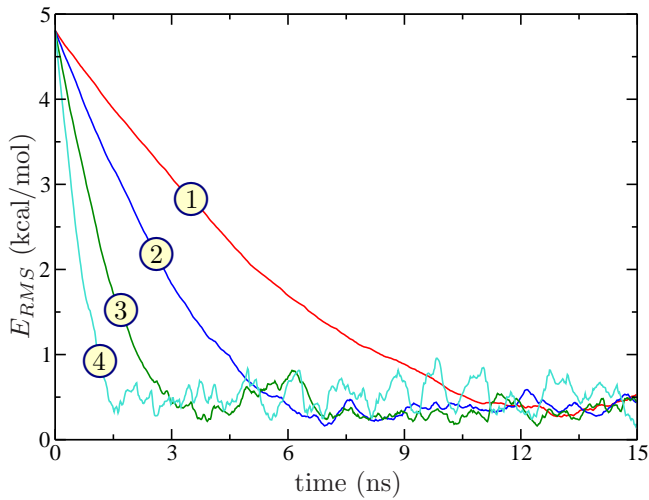}
\caption{\label{fig:5}%
RMS error of the free energy over $3.3\text{\AA}<R_g<6.3\text{\AA}$ at $T=300K$
for parallel tempering \texttt{ABMD} simulations using eight
replicas running at $T=300\,\mathrm{K}$,
$331\,\mathrm{K}$, $365\,\mathrm{K}$, $403\,\mathrm{K}$,
$445\,\mathrm{K}$, $492\,\mathrm{K}$, $543\,\mathrm{K}$ and
$600\,\mathrm{K}$ with $\tau_F = 90\,\mathrm{ps}$ (1),
$45\,\mathrm{ps}$ (2), $22.5\,\mathrm{ps}$ (3)
and $11.25\,\mathrm{ps}$ (4).
}
\end{figure}

In order to decrease the simulation time required for accurate free
energy estimates even further, the multiple-walker variation of \texttt{ABMD}
proves to be useful. For a moderate number of walkers, the speedup is nearly
linear, with an additional increase in accuracy coming from the better sampling of
the ``evolving'' canonical distribution (see Fig.\ref{fig:4}).
Parallel tempering improves both the speed and the accuracy even more.
To this end, we first ran \texttt{ABMD} with $\tau_F = 90\,\mathrm{ps}$
using 2, 4, 6 and 8 replicas at $T=300\,\mathrm{K}$,
$331\,\mathrm{K}$, $365\,\mathrm{K}$, $403\,\mathrm{K}$,
$445\,\mathrm{K}$, $492\,\mathrm{K}$, $543\,\mathrm{K}$ and
$600\,\mathrm{K}$ (during equilibrium \texttt{MD} runs, the peptide
configuration jumps between the two minima on a picosecond
timescale at $T=600\mathrm{K}$). In all cases, the $E_{RMS}$ was
found to be $\sim 1$ kcal/mol, or less as $t\to\infty$
(data not shown). Again,
the improvement in accuracy stems from the better sampling of the
``evolving'' canonical distribution.
Then, we ran 8 replicas with smaller values $\tau_F$
and were surprised that the accuracy does not degrade, even for
$\tau_F = 11.2\,\mathrm{ps}$ as shown in Fig.\ref{fig:5}.

\begin{figure}
\includegraphics[width=\linewidth,clip=true]{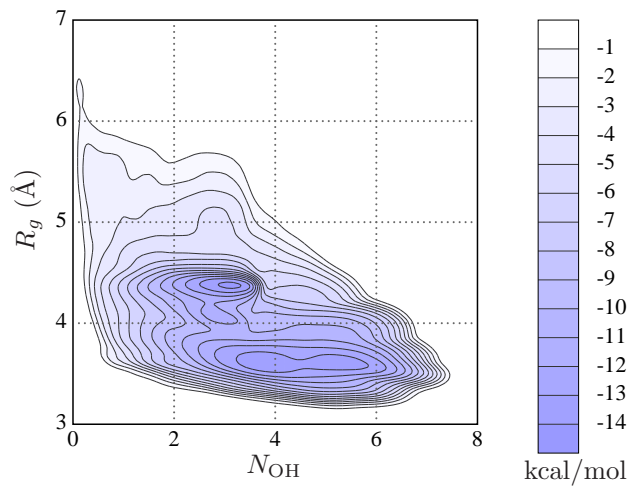}
\caption{\label{fig:6}%
Free energy map for \peptide peptide in the gas phase as a 
function of the collective variables $R_g$ and $N_{\mathrm{OH}}$.
(see Ref.\onlinecite{bicubic}
for details regarding the algorithm used to make this plot).
}
\end{figure}

Finally, we turn to aspects related to the \textit{general}
replica exchange method and illustrate its potential.
As already noted, by using replicas with different
collective variables and swapping these at prescribed rates, it is possible
to obtain projections of the free energy surface for the corresponding variables.
It is also possible to use previously obtained information with
respect to one collective variable to compute the free energy associated
with a different variable.
For example, suppose that instead of  the {\it one-dimensional} free energy profile
as as a function of $R_g$ already discussed, one realizes that what is actually
needed is a \textit{two-dimensional} profile that includes information 
 with respect to the number of O-H bonds along the backbone. The
two-dimensional free energy map is computationally quite expensive,
but the calculation can be greatly accelerated with the help of the general replica
exchange method.
We therefore simulated $8+1=9$ replicas. The eight replicas were run
at the previously stated temperatures, with each replica biased by a \textit{static}
(not evolving) biasing potential corresponding to the negated free energy associated
with  the radius of gyration $R_g$ at the corresponding temperature,
 as shown in Fig.\ref{fig:8}. The additional ninth replica was run
at $T=300\,\mathrm{K}$ with \texttt{ABMD}
flooding in the two collective variables, {\it i.e.,} $R_g$ and the number of
O-H bonds along the backbone as given by
\begin{equation}
  \nonumber
  N_{\mathrm{OH}} = \sum\limits_{\mathrm{O}, \mathrm{H}}
     \frac{1 - \big(\left.r_{\mathrm{OH}}\right/r_0\big)^{6\hphantom{1}}}
          {1 - \big(\left.r_{\mathrm{OH}}\right/r_0\big)^{12}}\,,
\end{equation}
where $r_{\mathrm{OH}}$ is the distance between a pair of hydrogen
and oxygen atoms and $r_0=2.5\text{\AA}$. The sum runs over the unique
O-H pairs (\textit{i.e.}, each O-H pair is counted only once),
with O and H separated by one or more amino-bases along the
backbone (27 pairs in total). In other words, we ``re-use'' the previously
computed free energies for $R_g$ to get the free energy in
the two-dimensional space $(R_g, N_{\mathrm{OH}})$: the eight first replicas
serve as a ``sampling enhancement device'' for the ninth replica. The
calculation is carried out in two stages: a ``coarse'' stage
($15\,\mathrm{ns}$ with $\tau_F = 10\,\mathrm{ps}$
and $4\Delta R_g = 0.25\text{\AA}$, $4\Delta N_{\mathrm{OH}} = 0.5$)
followed by a ``fine'' stage ($50\,\mathrm{ns}$ with
$\tau_F = 100\,\mathrm{ps}$ and $4\Delta R_g = 0.1\text{\AA}$,
$4\Delta N_{\mathrm{OH}} = 0.25$).
In both runs exchanges between four randomly chosen pairs of replicas
were attempted every $100\,\mathrm{fs}$. The final free energy map is shown
in Fig.\ref{fig:6}. It is clear that this two-dimensional free energy
landscape conveys additional information not contained in the one-dimensional
free energy plots already discussed.
In particular, it allows for a better characterization
of the ``globular'' states of the \peptide: specifically, it is
apparent from the Fig.\ref{fig:6} that there are at least
two such states with different values of $N_{\mathrm{OH}}$
(both correspond to the left minimum in Fig.\ref{fig:8}).
Of course,  one could have re-used the information in the one-dimensional
$R_g$ profiles to include other collective variables, in addition to
$N_{\mathrm{OH}}$.

The general replica exchange \texttt{ABMD} may also be advantageous
for explicit solvent simulations, which are often notoriously lengthy.
Specifically, if one is interested in the solute and the collective
variables do not depend on the solvent degrees of freedom, then the
number of replicas required to maintain an adequate exchange rate
depends only very weakly on the amount of solvent (which must of course
be sufficient as to adequately solvate the structure), provided that all
the replicas are simulated at the same temperature. This is because the
exchange probability does not explicitly depend on the potential energy
difference when the temperature of the replicas is the same. While 
not every choice of collective variables for different replicas will 
lead to decent exchange rates, one can nevertheless take advantage
of this property and use general replica exchange to enhance the
sampling in a solvated environment.

\begin{figure}
\includegraphics[width=\linewidth,clip=true]{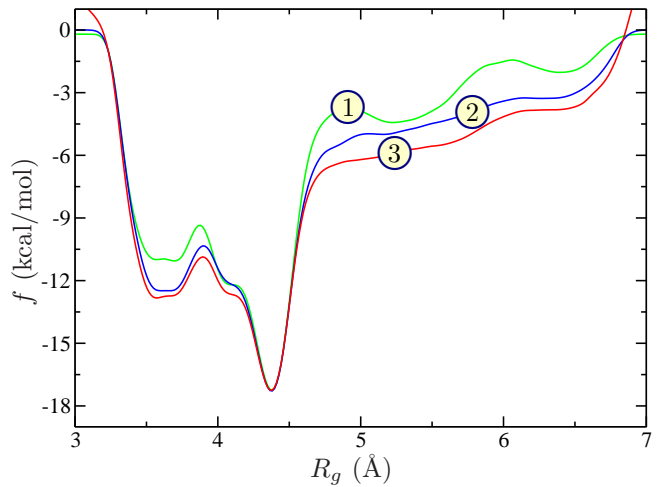}
\caption{\label{fig:7}%
Free energy for \peptide peptide solvated in cyclohexane
at $T=300\,\mathrm{K}$ as obtained via: coarse (non-equilibrium)
replica-exchange \texttt{ABMD} (1); finer (non-equilibrium)
replica-exchange \texttt{ABMD} (2); including the correction
coming from equilibrium biased replica exchange (3).
}
\end{figure}

In order to demonstrate the method in this regime,
we simulated the \peptide peptide at $T=300\,\mathrm{K}$ solvated
by 171 cyclohexane ($\mathrm{C}_6\mathrm{H}_{12}$) molecules
(the total number of atoms was 3139) under periodic boundary conditions
using the 
General\cite{GAFF} \texttt{AMBER} Force-Field (\texttt{GAFF}) for the solvent.
We used a \textit{truncated octahedron} cell of fixed size
(constant volume) that corresponds to the equilibrium
density at $T=300\,\mathrm{K}$ (the equilibrium density value was obtained from 
a $10\,\mathrm{ns}$ simulation  under constant pressure at $T=300\,\mathrm{K}$).
The Particle-Mesh Ewald (\texttt{PME}\cite{Darden_T_93}) method was
used for the electrostatic forces, with a $36\times 36\times 36$ \texttt{FFT}
grid and an $8\,\text{\AA}$ cutoff for the direct sum (same cutoff
was used for van der Waals interactions). First, we ran 10 replicas 
in the ``flooding'' mode ({\it i.e.},  under evolving biasing potentials) using
as collective variables the
distances $r_{\mathrm{CC}}$ between the backbone carbons separated by at least
2 amino-acids
(there are  10 such distances for
\peptide). We ran for $5\,\mathrm{ns}$ with $\tau_F=30\,\mathrm{ps}$
and $4\Delta r_{\mathrm{CC}} = 1\text{\AA}$, and then for
another $5\,\mathrm{ns}$
with $\tau_F=150\,\mathrm{ps}$ and $4\Delta r_{\mathrm{CC}} = 0.5\text{\AA}$,
attempting exchanges between five randomly selected pairs every $0.5\,\mathrm{ps}$.
As expected, at the start of the simulation, the
exchange rate was nearly 100\% decreasing later as the biasing potentials were
built up. By the end of the simulation, when all possible values of the
distances had been covered, the exchange rate was very disparate
between different pairs of replicas. However, for every
replica there was at least one other replica such that the exchange
rate between them was reasonable ({\it i.e.,} the whole simulation did not
degenerate into ten different non-interacting trajectories). We then set
$\tau_F=\infty$ in these 10 replicas, and added an eleventh replica whose
collective variable was chosen as the radius of gyration of the  heavy atoms.
In other words, as before, we use the ten replicas as a ``sampling enhancement
device'' for the last one. We then ran a two-stage flooding scheme:
a $5\,\mathrm{ns}$ coarser stage,
with $\tau_F=25\,\mathrm{ps}$ and $4\Delta R_g = 0.25\text{\AA}$;  followed
by a $10\,\mathrm{ns}$ finer stage, with $\tau_F=180\,\mathrm{ps}$ and
$4\Delta R_g = 0.2\text{\AA}$. As before, the exchange attempts between
5 randomly selected pairs of replicas were performed every $0.5\,\mathrm{ps}$.
The ``raw'' \texttt{ABMD}-computed free energy associated with $R_g$
after that stage is shown in Fig.\ref{fig:7}. In a next
step, we ran 64 biased simulations (each comprising of 11 replicas) for
$7\,\mathrm{ns}$ (first $2\,\mathrm{ns}$ for equilibration followed by
$5\,\mathrm{ns}$ of ``production'' runs) starting from different initial
configurations. We set $\tau_F=\infty$ in all replicas (static biasing
potentials) and recorded the values of $R_g$ in the eleventh replica every 10
picoseconds. We then used the log-spline algorithm of
Stone\cite{LogSpline} \textit{at. al.}
to estimate the (biased) log-density of the $R_g$ values at
\textit{equilibrium}. This led us to the final shape of the
free energy curve shown in Fig.\ref{fig:7}. 
Compared to the gas phase, the folded $\beta$-turn in the
cyclohexane solvated peptide
is clearly favored over the globular structure.
%
%
\section{\label{sec:outro}
Conclusions and outlook.
}
In summary, we have presented an \texttt{ABMD} method that computes
the free energy surface of a reaction coordinate using non-equilibrium dynamics. 
The method belongs to the general category of
umbrella sampling methods with an evolving potential, and is characterized
by only two control parameters (the flooding timescale and the kernel width)
and a favorable $O(t)$ scaling with molecular dynamics time t.
This scaling can be very important for large-scale 
classical \texttt{MD} biomolecular simulations when long
simulation times are required 
(see, for example Ref.\onlinecite{Shirts_M_2003},
and references therein).

\texttt{ABMD} has also been extended to support multiple walkers and
replica exchange. Both variations improve speed and accuracy of the
method due to the better sampling of the ``evolving'' canonical distribution.
The replica exchange \texttt{ABMD} has been generalized
to include different temperatures and/or collective variables, that move under
either an evolving or a static biasing potential. Aside from enhancing the sampling,
this swapping of replicas has several important practical advantages.
Most importantly, it enables one to obtain projections of the free energy
surface for {\it any number of collective variables} one might wish
to investigate. In addition, one can re-use previously obtained results
in order to enhance the sampling of new collective variables.
It is also possible to exploit the fact that exchange rates at the same temperature
are independent of
the potential energy to enhance sampling of a solute in a minimum
amount of solvent (for collective variables independent of solvent atom coordinates).
We have implemented the \texttt{ABMD} method in the \texttt{AMBER}
package\cite{Amber9}, and plan to distribute it freely. Here, we have
demonstrated the workings of the \texttt{ABMD} method with a study of the
folding of the \peptide peptide, The application of \texttt{ABMD} to
more complicated biomolecular systems is reserved for future publications.
%
%
\acknowledgments
This research was partly supported by NSF under grants ITR-0121361
and CAREER DMR-0348039.
In addition we thank NC State HPC for computational resources.
%
%
\appendix
\section{\label{ap:reference}
Reference free energy curve.
}
Here, we provide simulation details with regards to the reference free energy
curve.
We first ran short (5 ns, eight walkers with $\tau_F=60\mathrm{ps}$
and $4\Delta\xi=0.2${\AA}) multiple-walkers \texttt{ABMD}
at $T=600\,\mathrm{K}$ to reconstruct the global well. This was followed by
parallel tempering \texttt{ABMD} runs, using the biasing potential
obtained from the multiple-walkers simulation as the zero-time value
for the biasing potentials at different temperatures. We used eight replicas
at $T=300\,\mathrm{K}$,
$331\,\mathrm{K}$, $365\,\mathrm{K}$, $403\,\mathrm{K}$,
$445\,\mathrm{K}$, $492\,\mathrm{K}$, $543\,\mathrm{K}$ and
$600\,\mathrm{K}$ and attempted exchanges every 100 \texttt{MD} steps
($0.1\,\mathrm{ps}$).
The simulation started  with $\tau_F=60\mathrm{ps}$ and $4\Delta\xi=0.2${\AA},
and ran for $10^5$ exchanges. We then set $\tau_F$ to $600\mathrm{ps}$,
$4\Delta\xi$ to $0.1${\AA} and ran for $5\times 10^5$ more
exchanges. Finally, this was followed up with  $1\times 10^6$ more exchanges with
$\tau_F=6\mathrm{ns}$ and $4\Delta\xi = 0.1${\AA}.
\begin{figure}[t!]
\includegraphics[width=\linewidth,clip=true]{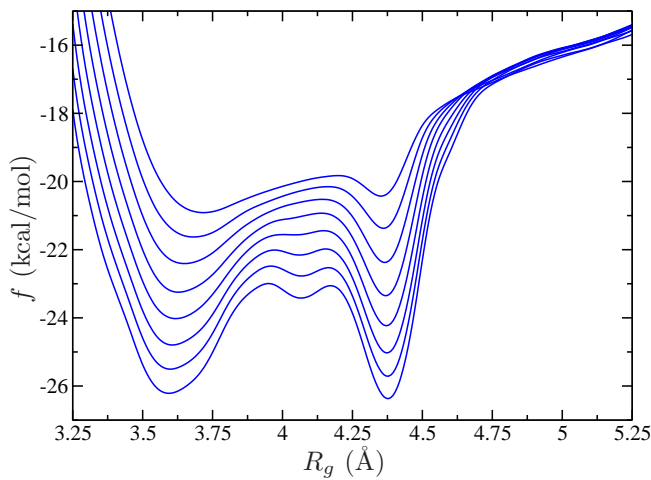}
\caption{\label{fig:8}%
The accurate free energies of \peptide peptide in gas phase
as function of $R_g$ at $T=300\,\mathrm{K}$,
$331\,\mathrm{K}$, $365\,\mathrm{K}$, $403\,\mathrm{K}$,
$445\,\mathrm{K}$, $492\,\mathrm{K}$, $543\,\mathrm{K}$ and
$600\,\mathrm{K}$ (from bottom to top).
}
\end{figure}

We then ran a very long ($3\times 10^7$
exchanges, $0.1\,\mathrm{ps}$ between exchanges) biased parallel
tempering simulation in the spirit of Ref.\onlinecite{Babin_V_2006}, in order
to get an \textit{a posteriori} error estimate. From the resulting
histogram it follows that the error does not exceed
$\approx 0.15\,\mathrm{kcal/mol}$ for 3.3{\AA}$<R_g<$6.3{\AA}.
The RMS error is probably much smaller, since 0.15 corresponds to
the absolute non-uniformity of the histogram, {\it i.e.}, the maximum error,
over 3.3{\AA}$<R_g<$6.3{\AA}. The accurate free energy curves as a
function of temperature are shown in Fig.\ref{fig:8}.
%
%

%
%
\end{document}